\documentclass[%
aps,pra,
superscriptaddress,
twocolumn
]{revtex4-1} 
\pdfoutput=1
\usepackage[colorlinks=true, linkcolor=blue, citecolor=blue, urlcolor=blue]{hyperref}
\usepackage{amsmath,amssymb,mathrsfs}
\usepackage{latexsym}
\usepackage{graphicx} 
\usepackage{epstopdf}
\usepackage{graphicx,epstopdf,color}
\usepackage{amsfonts}
\usepackage{bm}
\usepackage{bbm}
\usepackage{multirow}
\usepackage{natbib}
\usepackage{ifpdf}

\begin{document}

\preprint{}

\title{Underwater Acoustic Multiplexing Communication by Pentamode Metasurface}

\author{Zhaoyong Sun}
\email[These authors contribute equally.]{}
\affiliation{Key Laboratory of Noise and Vibration Research, Institute of Acoustics, Chinese Academy of Sciences, Beijing 100190, China}
\author{Yu Shi}
\email[These authors contribute equally.]{}
\affiliation{Department of Physics, Jishou University, Jishou 416000, Hunan, China}
\author{Xuecong Sun}
\affiliation{Key Laboratory of Noise and Vibration Research, Institute of Acoustics, Chinese Academy of Sciences, Beijing 100190, China}
\affiliation{University of Chinese Academy of Sciences, Beijing, 100049, China}
\author{Han Jia}
\email[Correspongding author:]{hjia@mail.ioa.ac.cn}

\affiliation{University of Chinese Academy of Sciences, Beijing, 100049, China}
\affiliation{ State Key Laboratory of Acoustics, Institute of Acoustics, Chinese Academy of Sciences, Beijing, 100190, China}
\author{Zhongkun Jin}
\affiliation{Key Laboratory of Noise and Vibration Research, Institute of Acoustics, Chinese Academy of Sciences, Beijing 100190, China}
\author{Ke Deng}
\email[Correspongding author:]{dengke@jsu.edu.cn }
\affiliation{Department of Physics, Jishou University, Jishou 416000, Hunan, China}
\author{Jun Yang}
\email[Correspongding author:]{jyang@mail.ioa.ac.cn}
\affiliation{Key Laboratory of Noise and Vibration Research, Institute of Acoustics, Chinese Academy of Sciences, Beijing 100190, China}
\affiliation{University of Chinese Academy of Sciences, Beijing, 100049, China}

\begin{abstract}
As the dominant information carrier in water, acoustic wave is widely used for underwater detection, communication and imaging.
Even though underwater acoustic communication has been greatly improved in the past decades, it still suffers from the slow transmission speed and low information capacity.
The recently developed acoustic orbital angular momentum (OAM) multiplexing communication promises a high efficiency, large capacity and fast transmission speed for acoustic communication.
However, the current works on OAM multiplexing communication mainly appears in airborne acoustics.
The application of acoustic OAM for underwater communication remains to be further explored and studied.
In this paper, an impedance matching pentamode demultiplexing metasurface is designed to realize multiplexing and demultiplexing in underwater acoustic communication.
The impedance matching of the metasurface ensures high transmission of the transmitted information.
The information encoded into two different OAM beams as two independent channels is numerically demonstrated by realizing real-time picture transfer.
The simulation shows the effectiveness of the system for underwater acoustic multiplexing  communication.
This work paves the way for experimental demonstration and practical application of OAM multiplexing for underwater acoustic communication.
\end{abstract} 
\maketitle

\section{Introduction}
The ocean, which accounts for 75\% of the earth's area, owns rich resources and mysterious species. 
Therefore, it is extremely important for human beings to understand, explore and develop the ocean.
Underwater communication technology is an important approach for ocean exploration.
However, light and electromagnetic waves are easy to decay in water so that they are difficult to be used for underwater detection.
The acoustic wave does not attenuate and is not scattered easily in water, thus it is the major carrier of underwater information and can be used to realize underwater communication and imaging.
There are problems of low frequency and low speed based on acoustic communication, which can not meet the growing demand of data transmission.
How to improve the transmission speed and expand the information capacity of acoustic communication is an urgent problem.
The acoustic orbital angular momentum (OAM) provides a feasible scheme.

The OAM is carried by the vortex wave, which has been widely studied and applied in the context of optics and electromagnetic wave\cite{sroor_high-purity_2020,wang_generating_2020,zhang_tunable_2020,ge_twisted_2020}.
The equiphase surface of vortex wave shows the shape of a spiral, where the rotation rate is mainly described by the OAM order (also named as topological charge).
This leads to a doughnut-shapped intensity profile.
These characteristics provide an additional freedom that makes vortex wave have more application prospects, for example, optical OAM beam has found significant applications from objects manipulation\cite{padgett_tweezers_2011} to imaging\cite{zhao_light_2020} 
and multiplexing communication\cite{gibson_free-space_2004,nagali_quantum_2009,wang_terabit_2012,bozinovic_terabit-scale_2013,wang_twisted_2019}.


Due to the similarity between optics and acoustics, OAM in acoustics has been investigated extensively in recent decades.
Many efforts on acoustic vortex wave have been reported, including OAM beam formed by active arrays\cite{hefner_acoustical_1999} and metamaterials\cite{jiang_convert_2016,jiang_broadband_2016,jiang_modulation_2020,
ye_making_2016,
fu_sound_2020},
OAM beam used for  objects manipulation\cite{zhang_acoustic_2020,sanchez-padilla_torsional_2020,marzo_holographic_2015,gong_particle_2019,zhang_reversals_2018,guo_manipulation_2020},
acoustic analogy of supper radiant of Kerr black hole\cite{faccio_superradiant_2019,cromb_amplification_2020} 
etc.

Besides, acoustic OAM used for communication is also one of the most  attractive topics.
The first attempt in acoustic OAM communication was realized by transducer array in air\cite{shi_high-speed_2017}.
Orthogonal acoustic vortex beam was created by the designed active array to provide more physical channels for information transmission.
An active sensor array was set on the receiving end to scan the acoustic field, and the information was demultiplexed by an inner product algorithm.
Thus, the transmission speed was limited by the sensor scanning and postprocess.
In order to overcome these issues, a passive, postprocess-free and sensor scanning-free OAM communication system was designed with airborne
 based demultiplexing metasurfaces (DMM)\cite{jiang_twisted_2018}, where the OAM could be lowered by $1$ after passing through a DMM.
In the system, the information was compiled into different OAM beams, where each beam was an independent channel. 
Considering the doughnut-shapped intensity profile of OAM beam and the fact that the spiral phase of $mth$ OAM beam can only be eliminated after passing through the $mth$ metasurface, the encoded information can be immediately decoded by the DMM with a microphone in the center behind each metasurface. 

Since the crucial role of acoustics in underwater circumstance, OAM multiplexing technology can be used to improve the efficiency of underwater communication.
Although there are some works to attempt to utilize OAM in airborne acoustic communication\cite{jiang_twisted_2018,shi_high-speed_2017,li_limits_2020,jiang_nonresonant_2020},
the application of OAM in underwater multiplexing communication remains to be further developed\cite{jiang_broadband_2016,zou_orbital_2020}.

One way to OAM multiplexing in underwater communication is to extend the demultiplexing metasurface system \cite{jiang_twisted_2018} to underwater case.
However, one of the biggest challenge for underwater metasurface is the difficulty of impedance matching. 
A suitable approach is using pentamode material (PM), which is fabricated from rigid solids (usually is used with metal) but exhibits fluid-like acoustic property\cite{milton_which_1995,cai_mechanical_2016,cai_phononic_2017,li_three-dimensional_2019,li_composite_2020,cai_effect_2020}.

In the 2-dimentional (2D) condition, PM is realized by hexagonal lattice structure.
The shear modulus of the latticed PM is small enough that it can be neglected and regarded as fluid in some frequency bands.
The effective acoustic properties can be effectively adjusted by tuning the geometric parameters, for example, it can be designed to have anisotropic or isotropic modulus\cite{kadic_practicability_2012,layman_highly_2013}.
This makes PM widely studied and applied in underwater acoustic wave control\cite{norris_acoustic_2009,chen_latticed_2015,chen_design_2016,chen_broadband_2017,chen_broadband_2019,tian_broadband_2015,su_broadband_2017,sun_design_2018,sun_quasi-isotropic_2019,lu_physically_2019,nie_scattering_2020,yu_latticed_2020}.

In this paper, we use 2D latticed PM to design an impedance matching underwater acoustic demultiplexing metasurface (PM-DMM) with thickness of 0.19$\lambda$ and radius of 0.44$\lambda$ ($\lambda$ denoted as sound wavelength).
With such a PM-DMM, a passive, postprocess-free and sensor scanning-free OAM-based underwater acoustic multiplexing communication system is constructed, which can decode data directly and promptly by using two transducers.
In the simulation, the information is encoded by mixing the plane wave and $1st$ order OAM beam together, where each of them carries a part of information.
Taking full advantage of the intensity profile of the OAM beam, the information carried by the $1st$ order beam can be detected with the probe placed at the back of the PM-DMM, while the other information in the plane wave can be detected with the probe located in front of the PM-DMM.
The simulation shows the effectiveness of the system for underwater acoustic multiplexing  communication.

\section{Metasurface For Vortex Wave}

\begin{figure*}[!hbtp] 
\includegraphics[width=6in]{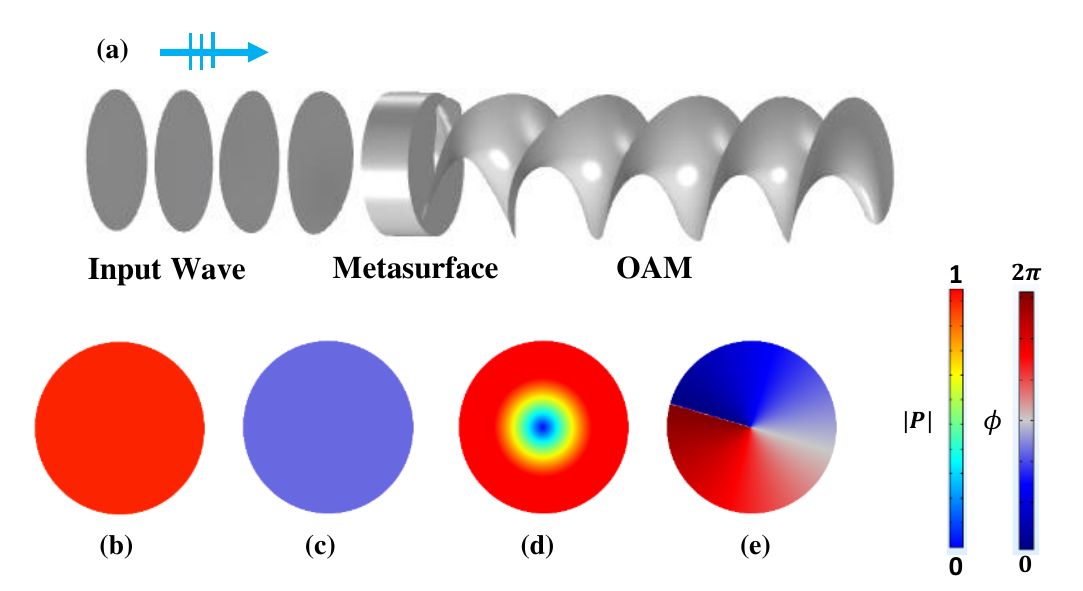}
	\caption{\label{fig:fig1} (Color online) Illustration of the OAM beam generated by a metasurface.  (a) The equiphase surface shows the working principle of the metasurface. 
	The amplitude (b) and phase (c) distribution of the plane wave at the cross-section in the waveguide.
	The doughnut shaped amplitude distribution (d) and the phase shift (e) of the $-1st$ order OAM beam at the cross-section in the waveguide.}
\end{figure*}

In the cylindrical coordinates system (denoted as {$r,\theta,z$ }), the pressure field of the $mth$ order acoustic OAM beam is expressed as:
\begin{equation}
	P=A_m(r,t)e^{i(kz+m\theta+\phi_m(t))},
	\label{eq:Pm}
\end{equation}
where $A_m (r,t)$ and $\phi_m(t)$ are the time dependent amplitude and phase, respectively.
And the index $m$, also named topological charge, indicates the order of the OAM.
From Eq.(\ref{eq:Pm}), it can be seen that the wavefront rotates around $z$ with the speed of $m\times 2\pi$ per revolution.
The sign of the topological charge indicates direction of wavefront rotation, i.e., a plus sign means rotating in anticlockwise direction, while the minus sign corresponds a clockwise rotation.
The beam with $m =0$ represents the plane wave, while the one with $m=\pm 1$ has a continue spiral shaped wavefront which leads to the zero pressure at the central axis.

Figure \ref{fig:fig1} shows the generation of a $-1st$ order beam by a $-1st$ order OAM metasurface.
In Fig.\ref{fig:fig1}(a), a plane wave  is emitted from left to the OAM metasurface.
Since the $-1st$ order OAM metasurface can lower the OAM of the incident wave, the transmitted wave is converted into a $-1st$ OAM beam.
The difference between the plane wave and the OAM beam can be clearly demonstrated by Fig.\ref{fig:fig1}(b-e).
The absolute pressure of the plane wave keeps constant in the $r-\theta$ plane(shown in Fig.\ref{fig:fig1}(b)), 
while that of the OAM beam exhibits a character of cylindrical symmetry with the value growing with $r$ from $0$ at $r=0$(shown in Fig.\ref{fig:fig1}(d)).
The phase of the plane wave also keeps constant in $r-\theta$ plane shown as Fig.\ref{fig:fig1}(c).
However, in OAM beam, the phase varies along $\theta$ direction as shown in Fig.\ref{fig:fig1}(e), which makes the equiphase surface like Archimedes Helicoid(see Fig.\ref{fig:fig1}(a)).
In the case of $|m|>1$, the wavefront is composed of $m$ intertwined helices.

From Eq.(\ref{eq:Pm}), it is easy to see that the vortex waves with different OAM orders are orthogonal\cite{li_limits_2020}, thus these OAM beams can be used as orthogonal channels to realize multiplexing communication.
The information capacity can be increased by $N$ times for the signal with $N$ vortex waves with different OAM orders.

Since a well designed OAM metasurface can increase or decrease the order of the vortex wave that passes through it\cite{jiang_twisted_2018},
OAM of a $mth$ order beam can be lowered to $m-n$ after $n$ identical $-1$ valued OAM metasurfaces.
This promises a metasurface based passive DMM system.
The mechanism is very intuitive: encoding the information into the vortex waves in advance and taking advantage of the null pressure at central axis,
one can obtain the amplitude and phase of the $nth$ order vortex wave by placing a detection point at the central axis behind the $nth$  metasurface.

The $mth$ order OAM metasurface of thickness $l$ should be designed by the fact that the phase difference at the two surfaces of the metasurface is linear correlation with $\theta$\cite{jiang_twisted_2018}:
\begin{equation}
	\varphi_{output}-\varphi_{input}=m \theta +\theta_0=kl,
	\label{eq:phase}
\end{equation}
where $k$ is the wave vector in the metasurface and $\theta_0$ is a constant.
Thus it is obvious that the velocity distribution $c(\theta)$ is:
\begin{equation}
	c(\theta)=\frac{2\pi fl}{m\theta+\theta_0}.
	\label{eq:c}
\end{equation}

From Eq.(\ref{eq:c}), it is easy to see that the refractive index of the metasurface varies gradiently with $\theta$.
Considering the impedance matching condition, the non-dimensional density and bulk modulus can be immediately derived as:
\begin{subequations}
  \begin{align}
	  \rho&=\frac{m\theta+\theta_0}{2\pi fl}c_0,\\
	  K&=\frac{2\pi fl}{(m\theta+\theta_0)c_0},
 \end{align}
\label{eq:rhoK}
\end{subequations}
where the dimensionless parameters are obtained through dividing the normal density and modulus by the corresponding values of background media.
And $c_0$ in Eq.(\ref{eq:rhoK}) is the acoustic velocity in the background media.
Such a metasurface can be used for passive multiplexing communication.



%
%
%
%

\section{The design of the unit cell }
\begin{figure}[!htbp]
	\includegraphics[width=3.25in]{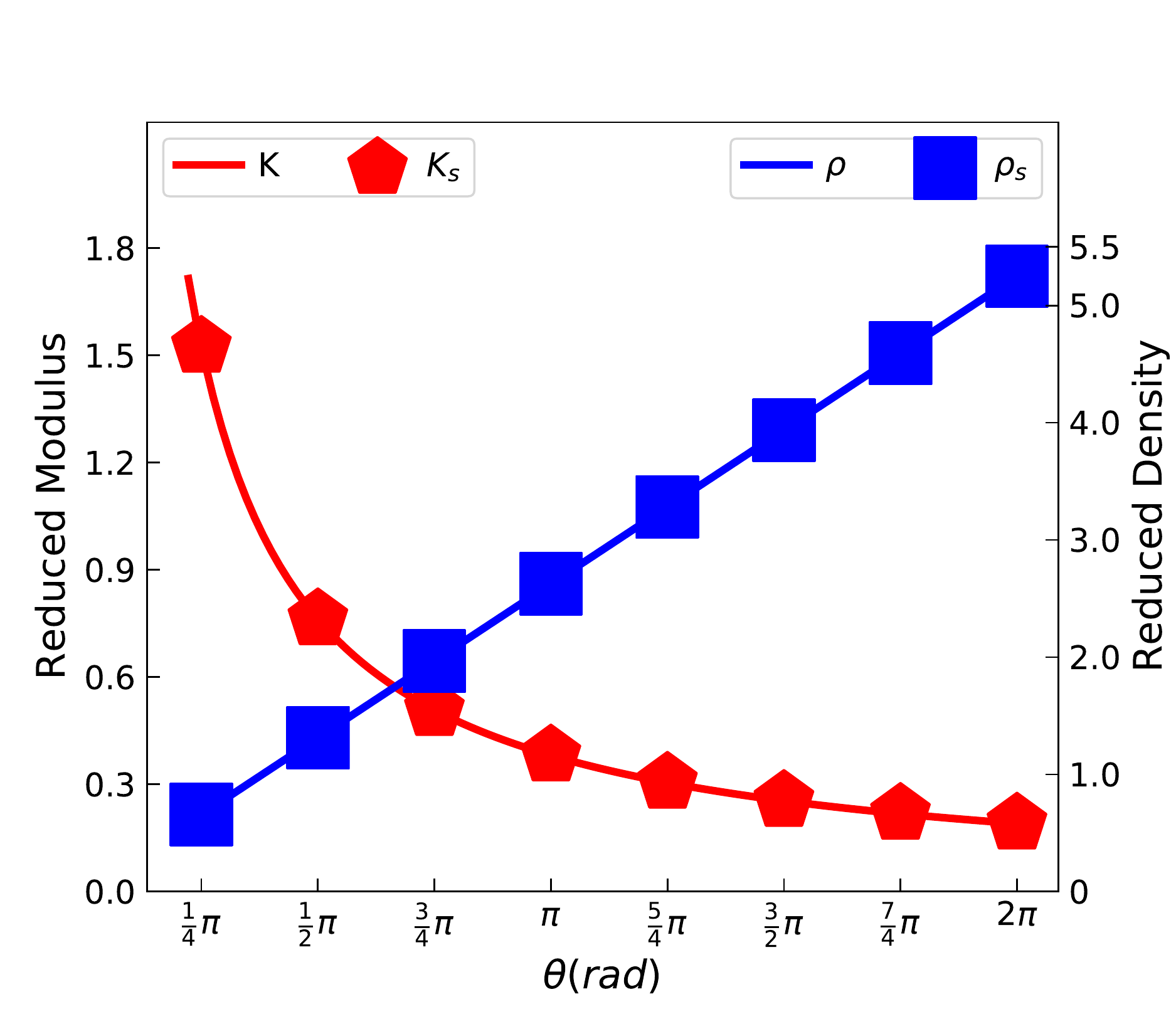}
	\caption{\label{fig:fig2}(Color online) The distributions of the bulk modulus and density of the PM-DMM. The profiles of continuously varying acoustic parameters(modulus $K$ marked by red solid line and density $\rho$ marked by blue solid line) and the corresponding discrete parameters(modulus $K_s$ by red pentagon and density $\rho_s$ by blue squares).} 
\end{figure}

%
%
%

The designed metasurface is a flat column with thickness $l=40$ mm and radius $r=92.38$ mm.
The distributions of $\rho$ and $K$ along $\theta$ direction are shown in Fig.\ref{fig:fig2}, marked by red and blue lines correspondingly.
\begin{figure*}[!htbp]

	\includegraphics[width=6in]{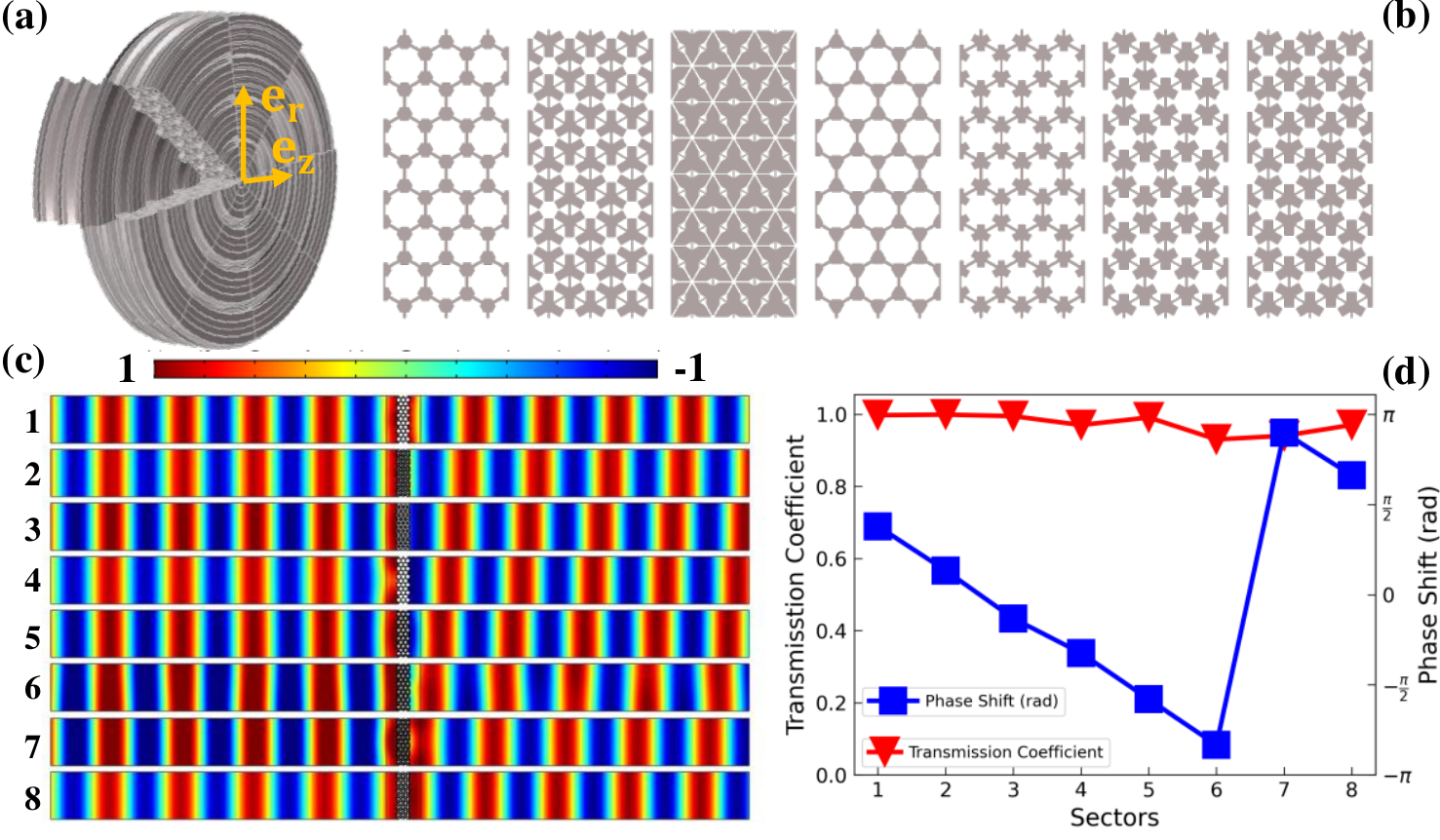}
	\caption{\label{fig:fig3}(Color online)
	(a) The PM-DMM constructed by 8 sectors.
	(b) The $r-z$ cross sections of the 8 sectors. 
	(c) Simulated sound transmission through the 8 cells.
	(d) The transmission coefficient (marked by blue triangles) and the phase shifts (marked by red squares) of the 8 cells.
	}
\end{figure*}
It's clear that the density increases with $\theta$ linearly, while the bulk modulus decreases inversely with $\theta$.
In order to realize the metasurface with metamaterials, the designed metasurface is pre-divided into $8$ equal sectors with central angle of $\pi/4$.
Then the continued parameters, density $\rho$ and bulk modulus $K$, are discretized into $8$ discrete groups which are marked by red squares (density) and blue triangles (modulus).

Since PM is excellent in underwater acoustic wave control\cite{chen_latticed_2015,chen_design_2016,chen_broadband_2017,chen_broadband_2019,tian_broadband_2015,su_broadband_2017,sun_design_2018,sun_quasi-isotropic_2019}, we use it to construct the designed metasurface.
The effect density and modulus of the PM structure are mainly determined by the geometry parameters of the unit cell.
According to the discrete parameters in Fig.\ref{fig:fig2}, the $8$ needed unitcells are obtained by retrieving the energy bands\cite{sun_design_2018,chen_latticed_2015}.
The basis material for the first three unitcells is aluminum with $E=69$ GPa, $\sigma=0.33$ and $\rho = 2700$ kg/m$^3$, 
while that for the last five unitcells is lead with $E=16.4$ GPa, $\sigma=0.44$ and $\rho = 11400$ kg/m$^3$.
Thereinto, $E, \sigma $ and $\rho$ represent Young's modulus, Poisson's ratio and density, respectively.
The background media is water with density of $\rho_0 = 1000$ kg/m$^3$ and bulk modulus of $K_0 = 2.25$ GPa.


By carefully tailoring, the $8$ rectangle-like structures are constructed and shown in Fig.\ref{fig:fig3}(b).
In order to evaluate the properties, the performance, i.e., transmission and phase shifts, of the $8$ structures are simulated by COMSOL Multiphysics.
The simulation is finished in a waveguide tube shown in Fig.\ref{fig:fig3}(c), where the waves at the frequency of $7100$ Hz are incident from left.
From Fig.\ref{fig:fig3}(c), it is clear that the transmitted phases of the $8$ structures vary gradiently.
A more accurate evaluation of the phase shift is shown in Fig.\ref{fig:fig3}(d) marked by blue squares.
The phase shift varies from $0.38\pi$ at sector $1$ to $-0.83\pi$ at sector $6$, jumps to $0.90\pi$ at sector $7$, and declines to $0.66\pi$ at sector $8$.
Apparently, the phase shift satisfies the relationship in Eq.(\ref{eq:phase}). 
The high transmission, marked by red triangles in Fig.\ref{fig:fig3}(c), implies good impedance matching performance of the structures.
The final PM-DMM, as shown in Fig.\ref{fig:fig3}(a), can be constructed by rotating the $8$ rectangle-like structures around the z-axis by $\pi/4$.

\begin{figure*}[!htbp]
	\includegraphics[width=4.5in]{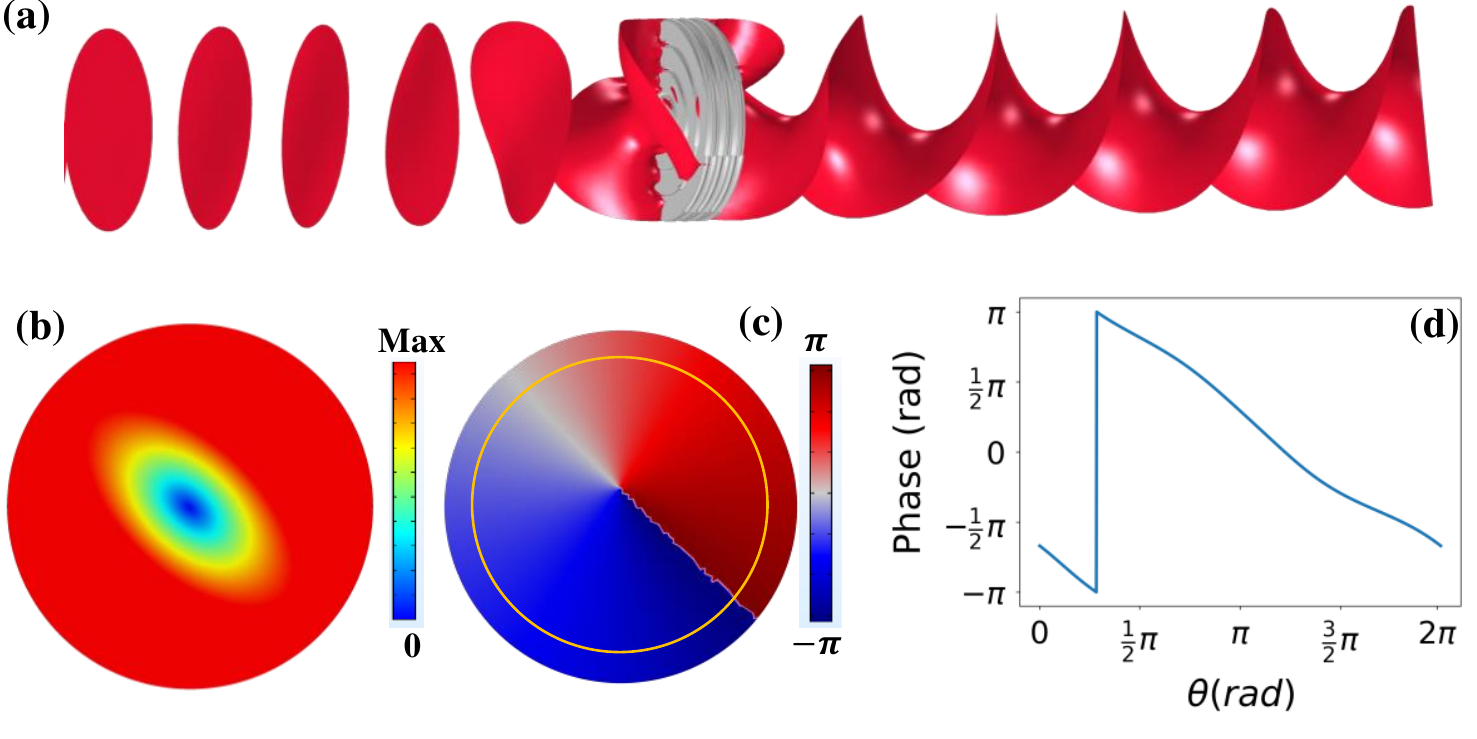}
	\caption{\label{fig:fig4}(Color online) 
	The simulated performance of the PM-DMM.
	(a) A plane wave at $7100$ Hz is converted to $-1st$ order OAM beam after passing through the PM-DMM. 
	(b) The pressure distribution of the $-1$st order OAM beam at $z=235 $ mm. 
	(c) The phase distribution of the $-1$st order OAM beam at $z=235$ mm. 
	(d) Phase shift along the yellow circle shown in (c) with $r=80$ mm.} 
\end{figure*}

\begin{figure*}[!htbp]
	\includegraphics[width=6.5in]{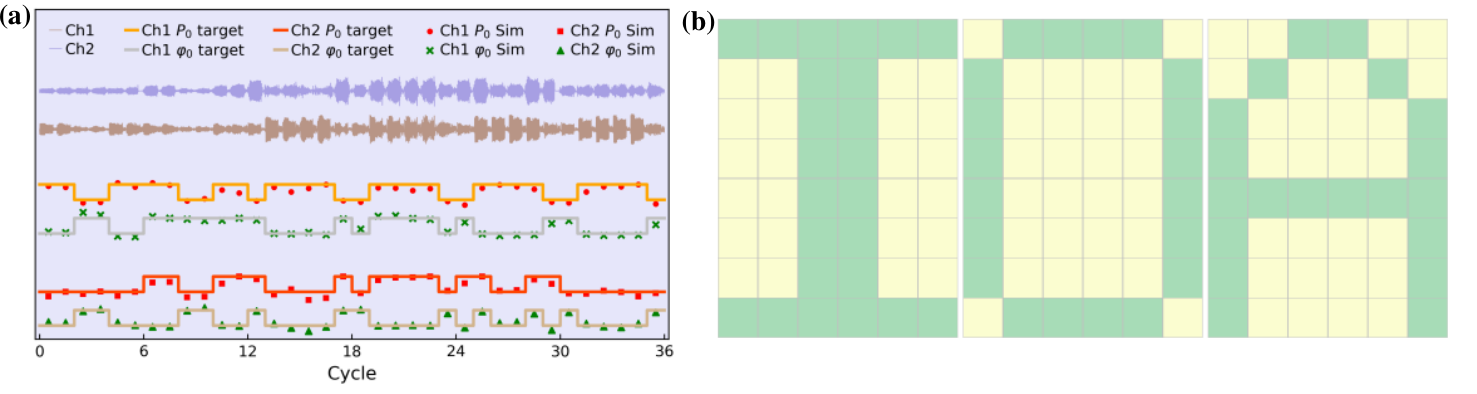}
	\caption{\label{fig:fig5}(Color online) Multiplexing communication with signal merged in plane wave (Ch1) and $1st$ order OAM beam (Ch2). (a) The received signals corresponding to the letters "IOA". 
	(b) Images independently retrieved from the received data carried by the two channels. } 
\end{figure*}

In order to verify the effectiveness of the OAM metasurface, the performance of the PM-DMM is simulated in a waveguide by the finite element solver COMSOL Multiphysics. 
In the simulation, the PM-DMM  is placed in the center of a cylindrical acoustic waveguide 
with a length of $1500$ mm and a radius of $92.38$ mm, which is exactly the same as the radius of the metasurface. 
The center point of the PM-DMM is set as the origin of the cylindrical coordinates.
A plane wave at the frequency of $7100$ Hz is emitted from the boundary at $z=-750$ mm to the metasurface.
The simulation results are shown in Fig.\ref{fig:fig4}.
In Fig.\ref{fig:fig4}(a), it can be seen that when the plane wave passes through the metasurface, 
the wave front changes from planes to anti-clockwise rotating spiral surfaces, which is the characteristics of the $- 1st$ OAM beam.
In addition, the amplitude and phase distributions of the $-1st$ order vortex wave at $z = 235$ mm are obtained and shown in Fig.\ref{fig:fig4}(b) and (c), respectively.
The amplitude shows doughnut distribution (Fig.\ref{fig:fig4}(b)), while the phase distribution is characterized by circumferential gradient (Fig.\ref{fig:fig4}(c)).
The yellow circle in Fig.\ref{fig:fig4}(c) is a cut line at $r=80$ mm, along which the phase distribution is extracted to exactly demonstrate the phase shift.
The phase shift along the cut line is shown in Fig.\ref{fig:fig4}(d).
Obviously, the phase shift varies gradiently between $-\pi$ and $\pi$ if the periodic mutation is ignored.
These results further confirm the OAM characteristics of the transmitted wave.
This proves that the designed PM-DMM can lower the OAM values of the incident wave by $1$ order.
The distributions of the amplitude and phase imply good performance of the metasurface. 
Thus it can be used for demultiplexing in underwater acoustic multiplexing communication.

\section{The Simulation Results}

%

We use the time domain simulation to demonstrate the multiplexing and demultiplexing of the PM-DMM in real-time underwater acoustic communication.
The multiplexing signal is a superposition of plane wave ($0th$ OAM beam) and $1st$ OAM beam, where the two waves are regarded as channel $1$ (Ch1) and channel $2$ (Ch2), respectively.
Two probes, denoted as probe $1$ and probe $2$, are set on the central axis in front of and behind the metasurface, respectively.
A detailed schematic image is shown in the supplementary material.
According to the analysis in the last section, the plane wave in the multiplexing signal will be converted to a $-1st$ OAM beam after passing through the PM-DMM, while the $1st$ OAM beam in the multiplexing signal will be converted to a $0th$ OAM beam (plane wave).
Consequently, the multiplexing signal can be demultiplexed by the metasurface.
Detailed examples of demultiplexing can be found in supplementary materials.


In this work, we  simultaneously use the binary amplitude shift keying (BASP) and binary phase shift keying (BPSK) formats to encode the information. 
The two different amplitudes in the multiplexing signal, i.e., normalized amplitudes of $1$ and $0.5$, encode the $0$ and $1$ of binary digits, respectively. 
In addition to amplitude modulation, phase modulation is also used to increase the information capacity, in which the $0$ and $1$ of binary are represented by the two different carrier phases that are $ \pi/2 $ degree apart from each others, respectively. 
In such a way, the data can be encoded into the two channels of the multiplexed signal to transmit information.
At the receiving end, the multiplexing signal can  be  demultiplexed  by  the  PM-DMM.  
Thus  the  data  encoded in different channels can be restored easily.

In the simulation of the acoustic OAM multiplexing communication, we encode every pixel in the images of letter "IOA" into the amplitude and phase of the waves.
The incident multiplexing signal is in pulse modulation denoted by $\Sigma _m p_0(T_0)e^{i(\omega t+m\theta+\phi_0(T_0))}$, in which $T_0$ is the period at $f_0=7100$ Hz. 
The pulse cycle in each channel is set as $20 T_0$ with duty ratio of $0.75$, where each cycle contains $2 bits$ data that correspond to two pixels of the images.
The amplitude and the phase in the incident signal are shown as the four rectangular lines in Fig.\ref{fig:fig5}(a) to be regarded as the objective data.
The orange and silver lines correspond to the initial amplitude and phase in Ch1, respectively, while the orangered and tan lines describe the amplitude and phase in Ch2 correspondingly.
The detected signals in the receiving end are shown as the saddle brown line (Ch1) and slate blue line (Ch2) in Fig.\ref{fig:fig5}(a), respectively. 
The distributions of the amplitude and phase in the two signals are obtain by Fourier transform and shown as the points with different symbols in Fig.\ref{fig:fig5}(a).
It can be seen that the received data is in good agreement with the objective data.
This is sufficient to show that the designed PM-DMM can separate the vortex waves with different OAMs to achieve the demultiplexing.
The received data shown in Fig.\ref{fig:fig5}(a) is transformed to binary format, where the binary information $0$ and $1$ represent two different colors of each pixel.
Thus the images are reconstructed as shown in Fig.\ref{fig:fig5}(b), which undistortedly form the pictures of letters "IOA".
The detailed process of the image reconstruction can be found in the supplementary materials.

%


\section{conclusion}

In conclusion, we have proposed and demonstrated a metasurface based passive demultiplexing system for underwater acoustic communication.
This system makes good use of the characteristic of null pressure at the central axis in the OAM beam to separate different OAM beams to demultiplex the signal.
This makes the system to be scanning free and post-processing free.
The metasurface, used for demultiplexing, is constructed by PM metafluid which is easy to be impedance matching with water, and is used for decoding with two probes before and after it.
In the simulation, $0th$ and $1st$ order OAM beams are used as two independent channels to carry data.
The simulation result shows the effectiveness of the system for underwater acoustic multiplexing  communication.
The mechanism of the demultiplexing system is not limited by numbers of the channels so that a multiplexing communication with more channels can be realized by adding more metasurfaces with equal numbers of transducers.
Thus the information capacity can be expanded to be large enough.
Moreover, the simulation in this paper provides a solid theoretical reference for the experimental exploration and practical application of underwater acoustic OAM multiplexing communication.

\begin{acknowledgments}
This work is partly surppoted by the National Natural Science Foundation of China (Grant No.11874383, 11964011, 11764016), the Youth Innovation Promotion Association CAS (Grant No. 2017029), the IACAS Young Elite Researcher Project (Grant No. QNYC201719) and Key-Area Research and Development Program of Guangdong Province (Grant No. 2020B010190002).
\end{acknowledgments}


\end{document}